# On the Influence of the FPGA Compiler Optimization Options on the Success of the Horizontal Attack

Ievgen Kabin[1], Alejandro Sosa[1], Zoya Dyka[1], Dan Klann[1] and Peter Langendoerfer[1,2]
[1]*IHP – Leibniz-Institut für innovative Mikroelektronik,* Frankfurt (Oder), Germany
[2]*BTU Cottbus-Senftenberg,* Cottbus, Germany
{kabin, sosa, dyka, klann, langendoerfer}@ihp-microelectronics.com

*Abstract*—This paper reports about the impact of compiler options on the resistance of cryptographic implementations against side channel analysis attacks. We evaluated four compiler option for six different FPGAs from Intel and Xilinx. In order to ensure fair assessment we synthesized always the same VHDL code, kept the measurement setup and statistical analysis method etc. constant. Our analysis clearly shows that the compiler options have an impact on the success of attacks but also that the impact is unpredictable not only between different FPGAs but also for an individual FPGA.

## I. INTRODUCTION

Many applications such as e-health, car-to-car communication, industry 4.0 etc. are requiring strong security means to ensure proper i.e. safe operation. In many cases it is not confidentiality what is needed but data integrity and authenticity. The latter two features can be provided using public key cryptography that is known to be time and energy hungry and due to that nor well suited for resource constraint devices as those used in in the Internet of things (IoT). ASICs supporting public key cryptography operations are an appropriate means to cope with limited resources of IoT devices. But the production of ASICs is pretty expensive when it comes to niche markets. For the latter FPGAs provide a reasonable alternative as they provide high performance at low energy cost at least compared to pure software solutions. In addition they can be reprogrammed which is especially for cryptographic algorithms an interesting feature.

When it comes to the implementation of cryptographic algorithms side channel analysis attacks (SCA) need to be taken into account. Otherwise successful SCA may render the use of cryptographic means void by extracting the keys. The latter allows then impersonation, falsifying digital signatures etc. During our earlier work we learnt that the compiler options used during synthesis have a significant impact on the SCA vulnerability of the resulting implementation even for ASICs [1]. This triggered the idea that similar might hold true for FPGA implementations as well. In order to evaluate this we decided to use our own design supporting the elliptic curve point multiplication that is vulnerable to horizontal differential side channel analysis attacks. The use of our own design has the benefit that we know exactly where the design is vulnerable. We used always the same VHDL code and our knowledge about its vulnerability for estimating the influence of the compiler options. We mapped the VHDL code on six different FPGAs using four different compiler options for each FPGA. So we got 24 different instantiations of the same VHDL implementation of our elliptic curve point multiplication. We analysed the vulnerabilities and differences in the vulnerabilities for all 24 instances in order to reveal the impact of the used compiler options. Our major findings are:

- Only for the Spartan 7 there is almost no difference in the attack success for all four investigated compiler option.
- For the other five FPGAs the compiler options have a significant impact on the attack success measured in the number of key candidates and their correctness. The impact of the compiler options also differs from FPGA to FPGA, i.e. no guideline can be given. So whenever implementing a design on an FPGA the influence of compiler options needs to be evaluated experimentally.
- The impact of the compiler options compared to their influence when synthesizing our design for an ASIC is negligible.

The rest of this paper is structured as follows. In the next section we shortly introduce our *kP* design. In section III the FPGAs and the compiler options used are discussed and the measurement setup is outlined. Our attack results, as well as their explanation, are provided in detail in section IV. Finally the paper finishes with short conclusions.

## II. OUR ECC DESIGN

### A. Implementation details

Our design is a hardware accelerator for the Elliptic Curve point multiplication for the NIST Elliptic Curve *B-233* i.e. it performs only a *kP* operation. The scalar *k* is an up to 233 bit long binary number and *P=(x, y)* is a point on EC *B-233*. The *kP* operation is the most time consuming operation for ECC. Nowadays ECC is applied for the exchange of shared secret keys, for mutual authentication of communication partners and for signing or verifying of messages. Corresponding to the ECDSA signature generation protocol the scalar *k* is a random number. This number that has to be kept secret, since otherwise the private key can be easily calculated [2]. In the EC authentication protocol the scalar *k* is the private key and has to be kept secret. Due to these facts we denote the scalar *k* further also as the *key*. The goal of attackers is to reveal the key i.e. the scalar *k*. The algorithm for the *kP* calculation has to be fast and resistant against different attacks, including SCA attacks. The Montgomery *kP* algorithm using Lopez-Dahab projective coordinates [3] is a bitwise processing of the scalar *k*. It is well-known and the most often implemented algorithm for the *kP* operation for ECs over $GF(2^n)$. This algorithm is fast and resistant against simple SCA attacks due to its regularity i.e. the operation sequence for processing a key bit in the main loop of the algorithm is independent of its value. Our implementation is based on Algorithm 2 presented in [4]. The processing of each key bit in the main loop of the algorithm requires 54 clock cycles. For the investigation reported here it is important that the design is vulnerable to horizontal differential SCA attacks. The implementation details of our *kP* design are published in [5]. The reason of the SCA leakage sources is the key dependant addressing of registers in the Montgomery *kP* algorithm. Possible countermeasures are proposed in [5]. The performed horizontal differential SCA attack is detailed described in [5]. The evaluation criterion of the performed attack is the correctness δ of the extracted key candidates. For each extracted key candidate we calculated its correctness as the relation of the number of the correct extracted bits to the length of the key. Thus, δ = 100% means that the key candidate is equal to real processed key i.e. the attack was 100% successful. δ = 0% means that no key bit was revealed correctly. This means that the inverted key candidate is the processed key with δ = 100%. Thus, δ = 50% is the worst case for the attacker. We synthesized our design using Vivado 2018.3 with the default synthesis and implementation options for the Spartan-7 FPGA and a clock frequency of 4 MHz. We captured an electromagnetic trace during a *kP* execution, see details for the measurement setup in section III.A. The blue line in Fig. 3 shows the calculated correctness δ of all 54 extracted key candidates. 19 key candidates have a correctness between 70% and 90%. The correctness of 4 key candidates is more than 90%. We selected our design with known for us vulnerability to investigate the impact of the FPGA's compiler options on the SCA resistance of the design.

### III. OUR EXPERIMENTS

Corresponding to [6] 50 percent of the FPGA market is covered with devices developed by Xilinx Inc., about 37 percent by Intel/Altera, the rest share Lattice Semiconductor, Microsemi, Texas Instruments and others. In our experiments we used the following boards from the two major manufacturers. All these boards are equipped with FPGAs manufactured by the Taiwan Semiconductor Manufacturing Company (TSMC). Details about used technology and process are given in TABLE I.

TABLE I.  FPGA MANUFACTURING DETAILS

|  | FPGA | TSMC's technology (Process) |
|---|---|---|
| **Intel** | MAX 10 FPGA Development Kit | 55 nm |
|  | Cyclone 10 LP FPGA Evaluation Kit | 60 nm (LP) |
|  | Cyclone V GX Starter Kit | 28nm (LP) |
|  | Arria V GX Starter Kit |  |
| **Xilinx** | Digilent Cmod S7 (Spartan 7) | 28nm (HPL) |
|  | Digilent Arty Z7-20 (Zynq 7020) |  |

FPGAs for the last two Intel boards and both Xilinx boards are produced in a 28 nm technology - the 28 LP (low-power) and the 28 HPL process (high-performance low-power) respectively. We ported our *kP* design to all 6 FPGAs listed in TABLE I. We synthesized and compiled always the same VHDL-code using the following tools:

- Quartus Prime v18.1 Standard Edition for Intel MAX 10; Cyclone 10 LP and Cyclone V GX FPGAs;
- Quartus II v15 for Intel Arria V;
- Xilinx Vivado 2018.3 WebPack for the Xilinx FPGAs.

For each FPGA we applied 4 different compiler options for the synthesis in order to evaluate the influence of the compiler settings on the resistance of the resulting designs against horizontal address bit DEMA attack. The applied options are:

- default flow;
- area optimization;
- performance optimization;
- power optimization.

Thus, we obtained 24 designs: 4 designs for each of the 6 FPGAs. In order to have equivalent settings for different tools and thus comparable implementations for the different optimizations strategies of the used compilers we used the mapping between different strategies as recommended in [7]. The applied options are shown in TABLE II. with their names as defined in [7].

TABLE II.  MAPPING BETWEEN QUARTUS OPTIMIZATION MODES AND VIVADO SYNTHESIS/IMPLEMENTATION STRATEGIES

| Quartus optimization mode | Vivado optimization strategy |
|---|---|
| Balanced (normal flow) | **Synthesis:** Vivado Synthesis Defaults **Implementation:** Vivado Implementation Defaults |
| Area (Aggressive) | **Synthesis:** Flow_AreaOptimized_high **Implementation:** Area Explore |
| Performance (high effort) | **Synthesis:** Flow_PerfOptimized_high **Implementation:** Performance Explore |
| Power (high effort) | **Synthesis:** Vivado Synthesis Defaults **Implementation:** Power_DefaultOpt |

First we determined is the maximum frequency for our *kP* design for each of the 6 FPGAs. Our results are:

- The maximal frequency for the implemented design is close to 120 MHz for all investigated FPGAs from Intel. For higher frequencies the synthesized designs did not meet the timing requirements and the calculated results were not correct.
- For the Xilinx FPGAs results were correctly calculated up to 200 MHz.

In order to allow fair comparison we determined a frequency of 100 MHz as the maximum operating frequency for all devices investigated. The parameters of the synthesized designs are given in TABLE III. and TABLE IV.

TABLE III. shows the resources used at Intel and Xilinx FPGAs, that are manufactured using the 28 nm technology. The mapping between resource names was done according to [7]. In the TABLE IV. we give information regarding the main resources for the Intel FPGAs manufactured in 55 nm and 60 nm technology respectively. The power estimation that is given in the tables was performed in vectorless mode after the

*route_design* stage. In contrast to the Quartus tools, the last versions of the Vivado Design Suite provide the power output values in Watts (earlier versions did that in *mW*), therefore it's hard to see the impact of selected optimization options on the estimated power consumption for boards equipped with Spartan-7 and Zynq-7020.

TABLE III. RESOURCES USED AT 28 NM FPGAS

| | Compiler optimization options | Combinational ALUT Usage for Logic/ LUT as logic | Dedicated Logic Registers/ Slice Registers | Power, mW |
|---|---|---|---|---|
| Spartan 7 | default | 5834 | 3706 | 70 |
| | area | 5833 | | 70 |
| | performance | 5834 | | 70 |
| | power | 5874 | | 68 |
| Zynq 7020 | default | 5833 | 3706 | 64 |
| | area | 5832 | | 63 |
| | performance | 5832 | | 64 |
| | power | 5875 | | 64 |
| Cyclone V | default | 5318 | 3842 | 64.56 |
| | area | 5130 | 3858 | 62.63 |
| | performance | 5341 | 3844 | 68.44 |
| | power | 5318 | 3892 | 64.02 |
| Arria V | default | 5338 | 3838 | 62.46 |
| | area | 5135 | 3851 | 59.65 |
| | performance | 5340 | 3908 | 63.25 |
| | power | 5338 | 3866 | 60.18 |

TABLE IV. RESOURCES USED AT INTEL 55 NM AND 60 NM FPGAS

| | Optimization option | Combinational with no register/with register | Register only | Power, mW |
|---|---|---|---|---|
| MAX10 | default | 5632/3058 | 662 | 200.41 |
| | area | 5573/3117 | 603 | 201.01 |
| | performance | 5585/2868 | 852 | 201.06 |
| | power | 5584/3106 | 614 | 186.11 |
| Cyclone 10 | default | 5734/2956 | 764 | * |
| | area | 5696/2994 | 726 | |
| | performance | 5717/2736 | 984 | |
| | power | 5636/3054 | 666 | |

*The generated power reports for Cyclone 10 contained only zeroes.

### A. Measurement Setup

We measured and analysed only the electromagnetic traces (EMTs) of the *kP* executions. The electromagnetic traces were captured during the *kP* operation using the near-field probe MFA-R 0.2-75 and recorded with a LeCroy HDO9404-MS oscilloscope. We placed the MFA-R probe close to one of the power decoupling capacitors of the core supply voltage $Vcc\_int$. We measured the traces with 40 GSamples/s which is the maximum sampling rate of the used oscilloscope. So we got 400 measured samples per clock cycle for the *kP* operations synthesized for and running at 100 MHz frequency. The duration of each *kP* execution is about 130 µs. Information about capacitor values is taken from the board schematics and listed in TABLE V. The placement of the EM probe during the measurements on each device is shown in Fig. 1. The position of the probe was fixed during all measurements of the same device i.e. all 4 designs with the different compiler options were ported and measured without any change in the measurement setup and probe placement.

TABLE V. CORE VOLTAGE POWER DECOUPLING CAPACITORS FOR BOARDS USED

| FPGA | Measurement place | Capacity |
|---|---|---|
| MAX10 | C219 | 1.0µF |
| Cyclone 10LP | C143 | 0.1µF |
| Cyclone V GX | C238 | 0.1µF |
| Arria V | C1143 | 0.1µF |
| Spartan7 | C40, C46 | 0.47µF, 47nF |
| Zynq 7020 | C125[a] | 0.47µF |

[a.] As the Digilent company doesn't provide a layout for Arty Z7-20 board we assume that it is the capacitor C125 from corresponding schematic due to its placement and physical dimensions.

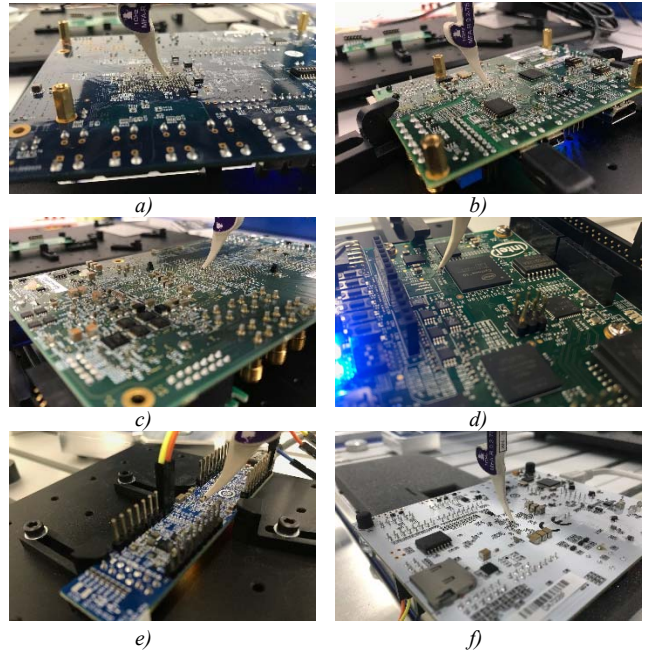

*a)* *b)* *c)* *d)* *e)* *f)*

Fig. 1. Devices under attack: boards equipped with Cyclone V GX (*a*), MAX10 (*b*), Arria V (*c*), Cyclone 10 LP (*d*), Spartan 7 (*e*), Zynq 7020 (*f*).

The shapes of the measured traces for the different boards differ significantly. For comparison Fig. 2 displays parts of the measured *kP* traces for each of the investigated boards.

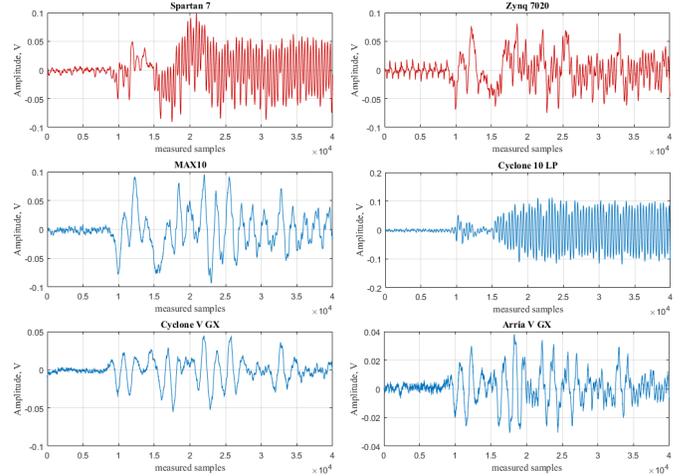

Fig. 2. Parts of the measured traces for designs with default setting, measured on all investigated boards.

Please note that in the *x* axis the same amount of measured samples is given for each of the traces. The amplitude of the measured signals i.e. the values on the *y* axis differ significantly for each trace. All traces shown in Fig. 2 are measured for designs synthesized with the default setting of the compiling tools. The two red traces were obtained using boards with Xilinx FPGAs. The blue traces were obtained by using boards with Intel/Altera FPGAs.

IV. ANALYSIS OF MEASURED TRACES

*A. Attack results depend on the frequency*

First we analysed the traces measured on the Spartan 7 FPGA with the goal to compare the attack results for two significantly different frequencies i.e. for 4 MHz (see section II) and for 100 MHz using the design synthesized with the compiler option "default". We performed the statistical analysis of the trace measured for the *kP* operation running at a frequency of 100 MHz using the comparison to the mean as described in [5]. As mention above, the placement of the EM probe was exactly the same for both measurements. The attack results for both experiments are shown in Fig. 3.

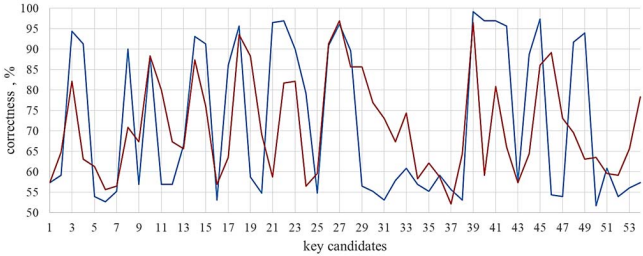

Fig. 3. Spartan 7 FPGA: results of the horizontal DEMA. The blue line corresponds to the design running at the 4 MHz and the red line corresponds to the design running at the 100 MHz.

There are 19 key candidates with a correctness of more than 90 percent for the design running at 4 MHz (see blue line in Fig. 3) and only 4 for the design running at 100 MHz (see blue line in Fig. 3). Please note that in addition to the different processing frequencies also the number of samples per clock cycle is different. For the 4 MHz design we got 625 samples per a clock cycle but only 400 samples per a clock cycle for the 100 MHz design. In our early experiments we investigated how reducing the sampling rate influences the correctness of the extracted key candidates for our *kP* design ported to a Xilinx Spartan 6 FPGA running at 4 MHz. The correctness of the extracted key candidates was almost the same if the number of the measured samples per a clock cycle was 100 or higher. If the trace was measured with a sampling rate resulting in less than 100 samples per clock cycle, the correctness of the extracted key candidates was reduced i.e. the attack was less successful. We don't evaluate this influence when attacking traces recorded for designs running at 100 MHz but we assume that it can be a one of reasons for the reduced success of the attack. It is important to note the fact that the correctness of the key candidates 7, 25, 34-36 and 52 is smaller than 60% for both frequencies. This was expected as the addressing of the design blocks in these clock cycles does not depend on the processed key bit value. The correctness of the key candidates 30, 31, 46 and 47 is higher than 75%. We did not expect such a high correctness for these key candidates due to the fact that the SCA leakage source based on the address bit phenomenon was not observable for these key candidates (see section II). We cannot explain these results analysing traces of the whole *kP* design. Understanding the reasons for this high correctness requires a detailed analysis of the electromagnetic traces (or power traces) of single blocks of our ECC design which is not possible for FPGAs.

*B. Attack results for the 100 MHz frequency*

The horizontal DEMA attack was performed for all 24 designs synthesized for a frequency of 100 MHz. For each attacked device 4 designs with the different compiler options were investigated. TABLE VI. shows the attack results, i.e. the number of key candidates with a correctness δ in selected intervals.

Selecting the compiler option "area" reduces the number of key candidates with a correctness higher than 90% from 12 to 3 for the Cyclone10 when we are considering just the numbers given in TABLE VI.

Results shown in TABLE VI. suggest also for all FPGAs except the Cyclone10 that there is almost no impact on the success rate. But the clock cycles in which the key candidates with a high correctness were extracted differ between the compiler options, i.e. the processes that are the high SCA leakage sources are different. The latter does not hold true for the Spartan 7. It shows the least influence of the compiler options as not only the number of key candidates with a certain correctness is almost constant, but that these key candidates are extracted in the same clock cycles.

TABLE VI. COMPARISON OF ATTACK SUCCESS

| FPGA | | number of key candidates with correctness δ | | |
|---|---|---|---|---|
| | | 50%≤δ<70% | 70%≤δ<90% | 90%≤δ≤100% |
| MAX10 | default | 31 | 15 | 8 |
| | area | 20 | 28 | 6 |
| | performance | 30 | 17 | 7 |
| | power | 28 | 20 | 6 |
| Cyclone 10LP | default | 24 | 18 | 12 |
| | area | 29 | 22 | 3 |
| | performance | 26 | 22 | 6 |
| | power | 27 | 19 | 8 |
| Cyclone V GX | default | 40 | 14 | 0 |
| | area | 40 | 14 | 0 |
| | performance | 44 | 10 | 0 |
| | power | 41 | 13 | 0 |
| Arria V | default | 40 | 12 | 2 |
| | area | 37 | 15 | 2 |
| | performance | 48 | 6 | 0 |
| | power | 38 | 14 | 2 |
| Spartan 7 | default | 31 | 19 | 4 |
| | area | 32 | 17 | 5 |
| | performance | 30 | 19 | 5 |
| | power | 31 | 19 | 4 |
| Zynq 7020 | default | 20 | 32 | 2 |
| | area | 27 | 25 | 2 |
| | performance | 25 | 26 | 3 |
| | power | 21 | 27 | 6 |

Due to the page limitation we don't represent the success of the performed attacks graphically.

Please note that the success of an electromagnetic analysis attack depends on the implemented design, the target FPGA, measurement set-up and the measurement position as well as on the applied statistical methods. For our experiments presented here we tried to keep all these factors constant. Please note that for the Xilinx FPGAs we used the same tool suite resulting in extremely different results. For the Spartan 7 there is almost no difference in the success rate per clock cycle for different compiler options whereas for the Zynq there are significant deviations. This is unexpected especially because both FPGAs are realized in the same TSMC technology and the resources used on the both FPGAs differ in a single LUTs only. Of course there are differences from FPGA to FPGA (manufacturing technology, geometry of the board etc.). In addition and eventually even more important due to the capacitors used on the different boards (see TABLE VI. ). As each of used boards has a different number of capacitors with different values, physical dimensions, placement on the boards, etc. i.e. there is a significant and device specific influence on the measured signal especially at high frequencies.

Due to the above mentioned facts we cannot explain why the compiler options lead to significant differences in the correctness of key candidates in different clock cycles. Such an explanation would require detailed knowledge about the compiler details in the sense of how the optimizations are realized. But we are convinced that even this knowledge would not allow to predict the attack success rate depending on the compiler option since the impact of the compiler options highly deviates for the two Xilinx FPGAs even though they are manufactured in the same TSMC technology. So, no guideline can be given i.e. whenever implementing a design on an FPGA the influence of compiler options needs to be evaluated experimentally.

V. CONCLUSIONS

In this paper we evaluated the impact of compiler options on the vulnerability of cryptographic designs against side channel analysis attacks. This idea was triggered by the fact that we noticed that the Synopsys compiler option "ultra" led to a significantly more resilient implementation of our design when we analysed it. As FPGAs are an interesting alternative to ASICs at least for niche markets, it is of some importance to know how the compile options and the target platform influence the vulnerability of an implementation. In order to ensure a fair comparison we kept all parameters that influence the success of an electromagnetic analysis attack such as the implemented design, the target FPGA, measurement set-up and the measurement position as well as on the applied statistical method constant. Our evaluation revealed that the impact of the compiler options for FPGAs is by far smaller than in the case of ASICs. We learnt in addition that even though we had a clear picture about the clock cycles in which our design is leaking information the compiler options may lead to a certain deviation in the sense that a high correctness of key candidates is achieved in unexpected clock cycles. We cannot explain this fact but are planning additional investigations such as analysing the influence of the frequency at which the design is running.